\documentclass[aps,prc,twocolumn,preprintnumbers,showpacs,superscriptaddress,amsmath,amssymb]{revtex4}
\usepackage{dcolumn}
\usepackage{bm}
\usepackage{amssymb}
\usepackage[german, english]{babel}
\usepackage{graphicx}
\usepackage{color}
\usepackage{textcomp}
\usepackage[letterpaper,total={7in,9.5in},top=0.75in,left=0.75in]{geometry}

\usepackage{array}

\begin{document}
\title{Towards an energy measurement of the internal conversion electron in the de-excitation of the Th-229 isomer}

\author{Simon Stellmer}\thanks{These two authors contributed equally.}
\affiliation{Atominstitut, TU Wien, 1020 Vienna, Austria}
\affiliation{Physikalisches Institut, Universit\"at Bonn, 53115 Bonn, Germany}
\author{Yudai Shigekawa}\thanks{These two authors contributed equally.}
\affiliation{Graduate School of Science, Osaka University, 1-1 Machikaneyama, Toyonaka, Osaka 560-0043, Japan}
\author{Veronika Rosecker}
\affiliation{Atominstitut, TU Wien, 1020 Vienna, Austria}
\author{Georgy A. Kazakov}
\affiliation{Wolfgang Pauli Institut, 1090 Vienna, Austria}
\affiliation{University of Vienna, 1010 Vienna, Austria}
\author{Yoshitaka Kasamatsu}
\affiliation{Graduate School of Science, Osaka University, 1-1 Machikaneyama, Toyonaka, Osaka 560-0043, Japan}
\author{Yuki Yasuda}
\affiliation{Graduate School of Science, Osaka University, 1-1 Machikaneyama, Toyonaka, Osaka 560-0043, Japan}
\author{Atsushi Shinohara}
\affiliation{Graduate School of Science, Osaka University, 1-1 Machikaneyama, Toyonaka, Osaka 560-0043, Japan}
\author{Thorsten Schumm}
\affiliation{Atominstitut, TU Wien, 1020 Vienna, Austria}

\date{\today}


\begin{abstract}
The first excited isomeric state of Th-229 has an exceptionally low energy of only a few eV and could form the gateway to high-precision laser spectroscopy of nuclei. The excitation energy of the isomeric state has been inferred from precision gamma spectroscopy, but its uncertainty is still too large to commence laser spectroscopy. Reducing this uncertainty is one of the most pressing challenges in the field.

Here we present an approach to infer the energy of the isomer from spectroscopy of the electron which is emitted when the isomer de-excites through internal conversion (IC). The experiment builds on U-233, which decays to Th-229 and populates the isomeric state with a 2\% fraction. A film of U-233 is covered by a stopping layer of few-nm thickness and placed between an alpha detector and an electron detector, such that the alpha particle and the IC electron can be detected in coincidence. Retarding field electrodes allow for an energy measurement. In the present design, the signal of the Th-229m IC electrons is masked by low-energy electrons emitted from the surface of the metallic stopping layer. We perform reference measurements with U-232 and U-234 to study systematic effects, and we study various means to reduce the background of low-energy electrons. Our study gives guidelines to the design of an experiment that is capable of detecting the IC electrons and measuring the isomer energy.
\end{abstract}

\maketitle

\section{Introduction}

The nucleus of the isotope Th-229 is unique in that it possesses an isomeric state of long half-life and very low energy in the range of a few eV. This low energy is comparable to the binding energy of valence electrons in neutral atoms, allowing to study the energy transfer between nuclei and valence electrons in the low-energy limit \cite{Mevergnies1969ceo,Mevergnies1972pot}. Moreover, the isomer energy corresponds to a wavelength of about 160\,nm, which is within the reach of current laser technology. The isomeric state in Th-229 has been suggested as the basis of a ``nuclear''clock \cite{Peik2003nls,Campbell2012sin,Peik2015ncb} and for precision measurements of possible drifts in fundamental constants \cite{Flambaum2006eeo,Rellergert2010cte}.

Direct information on the existence of the isomer \cite{Wense2016mta} and its half-life under IC \cite{Seiferle2017lmo} has been obtained from the detection of the internal conversion (IC) electron, detailed information on the excited nucleus has been derived from optical spectroscopy of the hyperfine structure \cite{Thielking2018lsc}, and determinations of the energy have been obtained from gamma spectroscopy \cite{Kroger1976fot,Reich1990eso,Helmer1994aes,Beck2007eso}. Direct optical excitation of the isomer has proven to be challenging \cite{Jeet2015roa,Yamaguchi2015esf,Wense2017ale,Stellmer2018oaa}, and further approaches such as excitation through a plasma \cite{Borisyuk2018eot} or by excitation of higher-lying nuclear states \cite{Yoshimi2018nrs} are being investigated. The isomer energy has been placed at 7.8(5)\,eV \cite{Beck2009ivf}. Theory estimations of the radiative half-life span a range between 1000 and 40\,000\,s \cite{Helmer1994aes,Ruchowska2006nso,Tkalya2015rla,Minkov2017rtp}. An improved uncertainty in the value of the isomer energy is urgently required for optical precision spectroscopy to commence.

In this work, we present a new and simple approach to access the isomer energy via spectroscopy of the IC electron \cite{Seiferle2017fso}. A thin layer of U-233 is covered by a metallic stopping layer (\textit{e.g.}~gold) with a thickness matched to the range of Th-229(m) recoil nuclei originating from the U-233 alpha decay. The Th-229m recoil nucleus undergoes IC with a half-life of $7(1)\,\mu$s \cite{Seiferle2017lmo}, and the emerging electron may leave the stopping layer and reach an electron detector. With retarding field grids placed between the sample and the electron detector, the isomer energy can be determined as the sum of the work function of the retarding field grid plus the value of the retarding voltage at which the signal vanishes \cite{Seiferle2017fso}. In our approach, the thin U-233 layer is applied directly onto an alpha particle detector:~in this way, U-233 alpha decays and IC electrons are measured in coincidence. This gating scheme holds the promise to enable a signal-to-noise ratio sufficient for the detection of the IC electron and for a measurement of its energy.

This approach builds on several advantages. At first, the alpha decay of  U-233 recoil is a well-established source of Th-229m recoil nuclei, where a 2\% branching into the isomer has recently been verified experimentally \cite{Thielking2018lsc}. Second, it seems conceptually simpler to use IC as a resource \cite{Wense2016mta,Seiferle2017lmo,Wense2017ale} rather than to suppress IC and search for the radiative decay \cite{Jeet2015roa,Yamaguchi2015esf,Stellmer2016fso}.

The isomer's half-life under IC depends on the electronegativity of the surrounding material, an effect that has been studied intensively with U-235m \cite{Mevergnies1969ceo,Mevergnies1972pot} and Th-229m \cite{Seiferle2017lmo,Seiferle2017privatecomm}. Within certain limits, the half-life can thus be tuned by choice of the stopping layer material. In presence of prompt electrons originating from the U-233 alpha decay within the first $\mu$s, one might chose a metal with high electronegativity such as to delay the IC electron signal. Once the Th-229m IC half-life has been measured in a few different materials with well-known electronegativity, the IC half-life can be derived for any other material. This is in stark contrast to searches for the VUV gamma ray, as the radiative lifetime is not yet known and a large parameter range has to be considered.

This approach requires a stopping layer material in which the 84-keV recoil ions are stopped, yet the few-eV electrons can exit the layer. Indeed, typical metals have similar ranges for 84-keV recoil ions and few-eV electrons \cite{Xraybooklet,Crowell67}.

In our approach, the IC process takes place in the bulk of the stopping layer, as opposed to a range of experiments where nuclei in the isomeric state are deposited onto a surface \cite{Wense2016mta}. While surfaces may be prone to chemical contaminations on the microscopic scale, the bulk material  is comprised of a well-controlled material.

Radioactive decay in conductive materials is known to produce prompt low-energy electrons, which may mask the sought-after IC electron signal. Reference measurements with U-234 and U-232 will allow for a subtraction of the prompt electron signal. As we will show below, the expected signal amplitude can be calculated with high fidelity, allowing for consistency checks of the data.

\section{Simulations and design of experiment}

\begin{figure}
\includegraphics[width=\columnwidth]{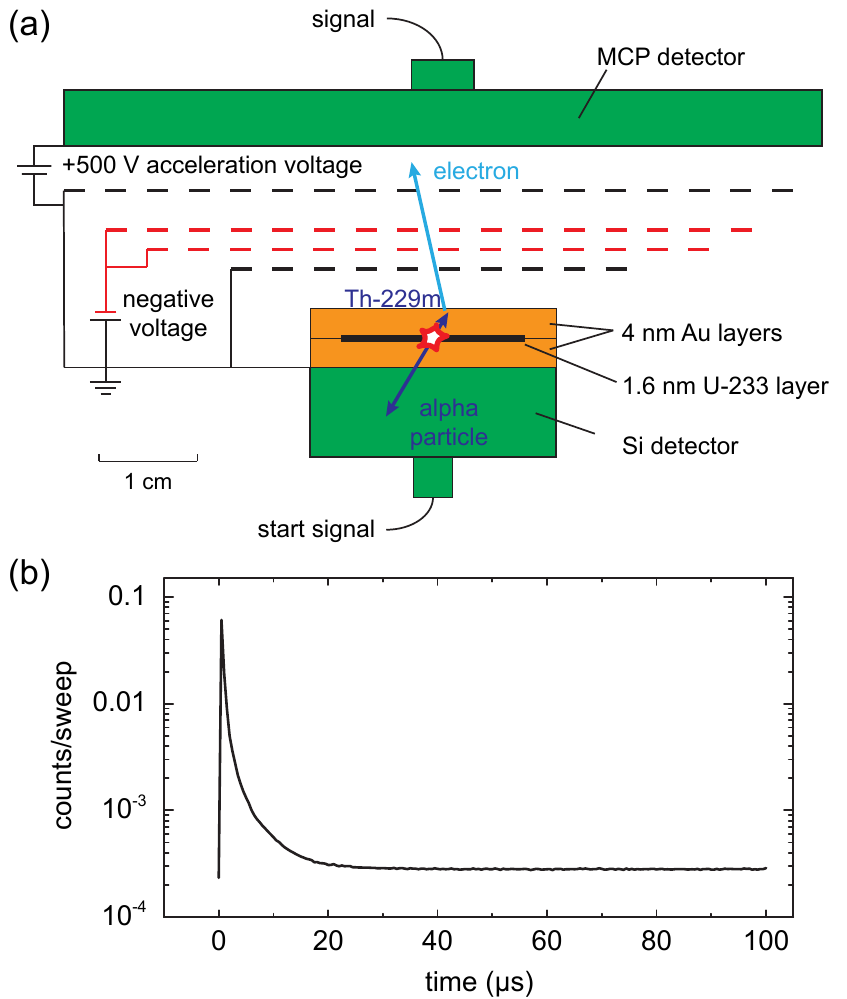}
\caption{(a) Experimental setup: A thin uranium layer is sandwiched in-between two gold layers and placed between a Si detector (alpha particles) and an MCP detector (electrons). (b) A typical time trace of the MCP signal, triggered by an alpha decay that is registered by the Si detector. The counts are integrated over bins of 500\,ns length and normalized to the number of detected alpha decays (``sweeps''). This data set represents an average over $5\times 10^7$ sweeps, it is taken with the U-233 sample at zero retarding field.}
\label{fig:fig1}
\end{figure}

A thin layer of U-233 is sandwiched between two stopping layers and placed on a Si detector. The alpha particle produced in the decay $^{233}{\rm U}\rightarrow {}^{229}{\rm Th(m)}$ is registered by a Si detector, and the Th-229m nucleus recoiling out of the U-233 layer is slowed down in the stopping layer. It de-excites via IC under the emission of an electron, which is detected by a microchannel plate (MCP) detector. A variable retarding voltage applied to metal meshes placed between the sample and the MCP detector allows for energy-selective detection. The experimental setup is depicted in Fig.~\ref{fig:fig1}(a) and described in more detail in Sec.~\ref{sec:expsetup}.

\subsection{Choice of the stopping layer material}

We choose gold as the stopping layer material for two reasons: at first, the electronegativity of Au is one of the largest among all metals, which extends the IC half-life. Secondly, the IC half-life of Th-229m on Au has recently been measured \cite{Seiferle2017privatecomm}. The value for the half-life turned out to be $t_{1/2}=7(1)\,\mu$s (lifetime $\tau = 10(1)\,\mu$s), such that this value can be considered a fixed parameter in the search for the signal.

\subsection{Numerical simulations to optimize the stopping layer thickness}

\begin{figure}
\centering
\includegraphics[width=\columnwidth]{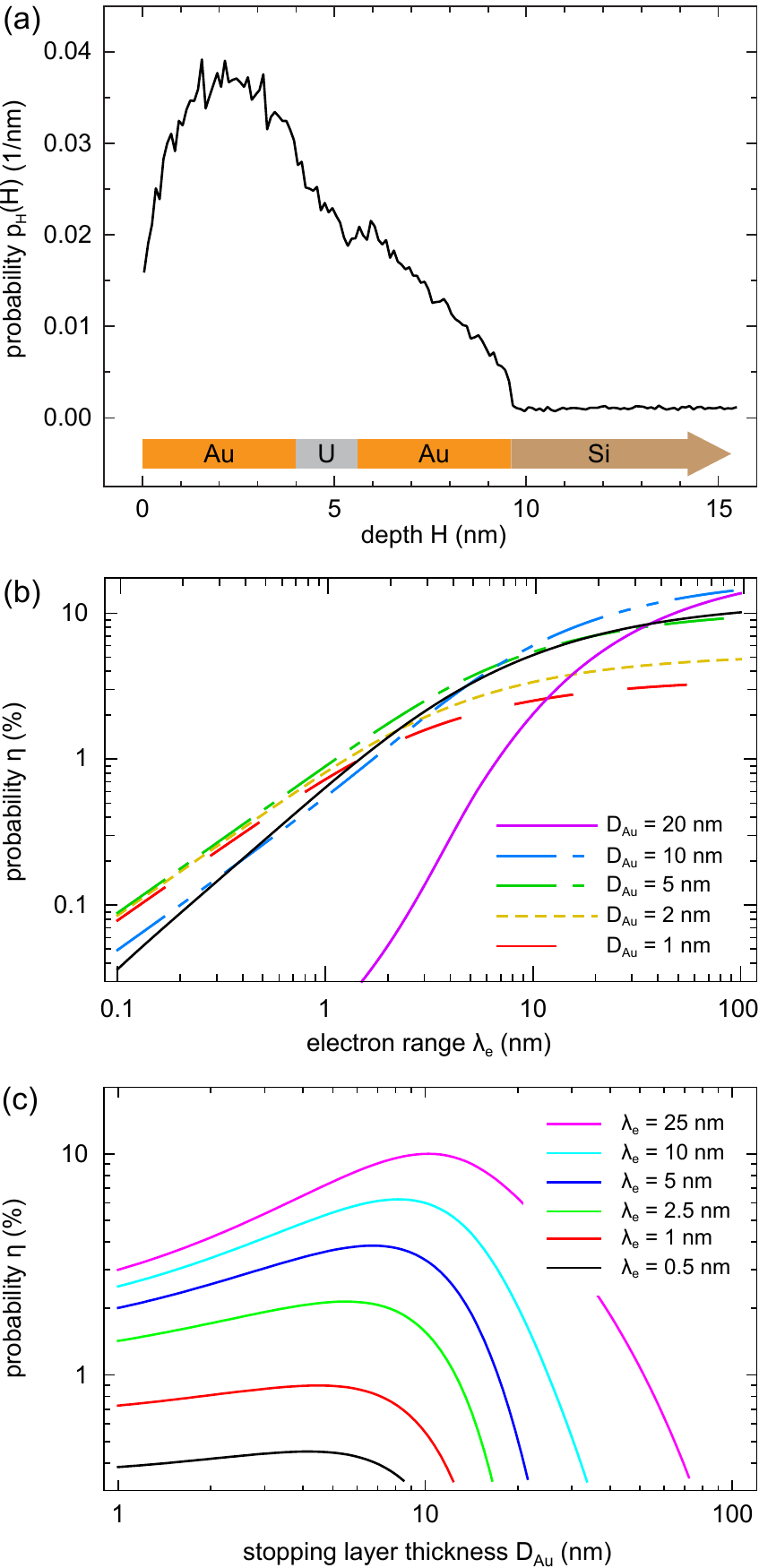}
\caption{Simulation of the experiment. (a) Probability distribution $p_H(H)$ for a Th-229 recoil nucleus to be emitted into the upper hemisphere and stopped at a depth $H$ from the outer surface, simulated with the help of the SRIM package for the configuration presented in Sec.~\ref{sec:expsetup}. (b) Probability $\eta$ that the decay ${\rm ^{233}U} \rightarrow {\rm ^{229m}Th}+\alpha$ is accompanied by the emission of an $\alpha$-particle into the lower hemisphere, the recoil nucleus is stopped within the source, and the conversion electron leaves the source towards the upper hemisphere, as a function of the inelastic mean free path $\lambda_e$ of the electron. The solid black curve corresponds to the calculation according to Eq.~(\ref{eq:eta}) for the SRIM-simulated distribution of recoil nuclei presented in (a), and styled colored curves correspond to approximate estimations of recoil nuclei (see Appendix~\ref{app:I}) for different thicknesses $D_{\mathrm{Au}}$ of the stopping layer. (c) Probability $\eta$ in dependence of the stopping layer thickness.}
\label{fig:fig2}
\end{figure}

For an estimation of the distribution of Th-229 recoil nuclei, we use the SRIM-simulation package \cite{SRIM}. As it has been shown in Ref.~\cite{Wense2018otd}, SRIM simulations significantly underestimate the stopping ranges of recoil nuclei in polycrystalline structures (\textit{e.g.}~produced by an evaporative technique), but slightly overestimate these ranges in amorphous structures obtained by electrodeposition techniques. For preliminarly estimations, we use a semi-analytical theory (see Appendix~\ref{app:I}), where the stopping lengths and stragglings have been taken from the SRIM package \cite{SRIM}. Also, for the specific geometry used in this study (1.6\,nm of uranium sandwiched between two 4-nm gold layers deposited onto the 50-nm thick Si layer of the detector), we perform numerical simulations of transport of the recoil ions (TRIM module of the SRIM package). This simulation is run for $N_{\rm sim}=3\times 10^5$ recoil ions, whose initial positions are randomly distributed within the uranium layer, and whose initial velocities are randomly directed into the upper hemisphere (corresponding to alpha particles moving towards the Si detector). Kinetic energies of all recoil ions are taken to be 84~keV. To estimate the probability density $p_H(H)$ that the alpha particle will be emitted into the lower hemisphere, and the recoil nucleus will be stopped at a depth $H$ from the outer surface, we select a number of bins with mean depths $H_n$ and thicknesses $\Delta H=0.1$~nm. Then,
\begin{equation}
p_H(H_n)\approx\frac{\Delta N}{\Delta H}\frac{1}{2 N_{\rm sim}}.
\label{eq:distr_d}
\end{equation}
Here, the factor of 2 in the denominator describes the fact that only half of the $\alpha$-particles are emitted into the lower hemisphere. The simulated distribution of $p_H$ is presented in Fig.~\ref{fig:fig2}(a).

Th-229m recoil nuclei born in the isomeric state de-excite primarily via the conversion mechanism. Conversion electrons will have an energy of a few eV above the Fermi level. Inelastic mean free paths $\lambda_{e}$ of electrons in metals have minima of 0.5 to 1\,nm at about 20 to 200\,eV \cite{Xraybooklet}, and grow to tens of nm (in gold) when the energy drops to about 1\,eV \cite{Crowell67}. For a rough estimation of the escape probability, we will neglect the elastic scattering. Then, the probability $P_{\rm e, out}(H_n, \lambda_e)$ of the electron born at a  depth $H_n$ from the surface to escape the source without inelastic scattering is
\begin{equation}
\begin{split}
P_{\rm e, out}(H_n, \lambda_e) =& \frac{1}{2} \int\limits_{0}^{\pi/2}
   \exp\left[-\frac{H_n}{\lambda_e \cos(\theta)}\right]
   \sin(\theta) d\theta \\
=&\frac{H_n}{2 \lambda_e} \left( \exp\left[-\frac{H_n}{\lambda_e}\right] 
  - \Gamma\left[0,\frac{H_n}{\lambda_e}\right] \right),
\end{split}
\label{eq:proboute}
\end{equation}
where $\Gamma(x,y)=\int_y^\infty t^{x-1}e^{-t}dt$ is the incomplete gamma function. The probability $\eta$ that the decay ${\rm ^{233}U} \rightarrow {\rm ^{229m}Th}+\alpha$ will be accompanied by emission of the $\alpha$-particle into the lower hemisphere, the recoil nucleus will be stopped within the source, and the conversion electron will leave this source towards the upper hemisphere, may be estimated as
\begin{equation}
\eta=\int P_{\rm e, out}(H, \lambda_e)\, p_H(H) \,dH.
\label{eq:eta}
\end{equation}
The probability $\eta$ is plotted in Fig.~\ref{fig:fig2}(b) for various values of the gold layer thickness $D_{\mathrm{Au}}$. For an estimated mean free path $\lambda_e$ of the IC electron of 2.5\,nm and a sample geometry of 1.6\,nm uranium and 4\,nm gold, we arrive at a probability $\eta = 2\%$. This value changes only very mildly for variations of $D_{\mathrm{Au}}$ between 1 and 10\,nm. This indicates that the experiment is tolerant against experimental inhomogeneities of the layer thickness. Figure~\ref{fig:fig2}(c) shows the same simulation, but with the stopping layer thickness $D_{\mathrm{Au}}$ varied continuously for various values of the electron range $\lambda_e$.

\subsection{Expected signal and background}

We expect the MCP signal to feature three components: a strong prompt signal in coincidence with the alpha decay, the sought-after IC electron signal itself, and a uniform background. Secondary electrons emitted from the material surface are not considered in this subsection.

\emph{Prompt signal ---} The prompt signal coincides with the U-233 alpha decay and may be caused by the recoil ion penetrating the stopping layer, gamma rays and X-rays, high-energy conversion electrons, as well as secondary photons and electrons. This signal is expected to decay on very fast timescales, certainly faster than $1\,\mu$s, and will not interfere with the IC electron signal.

\emph{IC electron signal ---} The IC electron count rate $C_{IC}$ per alpha decay can simply be expressed as
\begin{equation}
C_{IC}(t) = b \cdot \eta \cdot \lambda \cdot \epsilon_{e^-} \cdot \frac{1}{\tau}\ e^{-t/\tau},
\label{eqn:signal}
\end{equation}
where $t$ denotes the time after the alpha decay. All parameters are either pre-defined or can only be varied within very limited bounds. The branching ratio $b$ of the U-233 parent nucleus to populate the isomer has been estimated \cite{Kroger1976fot} and experimentally verified to $b = 2\%$ \cite{Thielking2018lsc}. The combined probability that the alpha particle will be registered by the Si detector, the Th-229 recoil nucleus will be stopped within the source, and the IC electron will leave the layer towards the MCP hemisphere, has been simulated above, with $\eta = 2\%$ being a reasonable estimate. The parameter $\lambda$ denotes the geometric transmission probability of electrons through the retarding field grids (see below), which is $\lambda \approx 40\%$ for our setup. The MCP detection efficiency for electrons $\epsilon_{e^-}$ accelerated to 500\,eV is typically 50\%.

In the experiment, we will perform a time binning into bins of reasonable length, \textit{e.g.}~$\Delta t = 0.5\,\mu{\rm s} \approx \tau/20$. In this discretized fashion, the time-binned IC electron counts per bin $C'_{IC}$ become $C'_{IC} = C_{IC} \cdot \Delta t$.

\emph{Background ---} The radioactivity of the sample creates a purely Poissonian background without time dependence. The background counts per time bin in a single sweep (\textit{i.e.}, a single time trace initiated by an $\alpha$-decay) are
\begin{equation}
C'_{bg} = c \cdot A \cdot \Delta t,
\label{eqn:background}
\end{equation}
where $A$ is the activity of the sample. The coefficient $c$ denotes the average number of MCP counts induced by an alpha decay. The alpha decay is a very fast process, and we indroduce an artificial electronic dead time of 500\,ns on the MCP signal after every count. In this way, at most one MCP count is registered. As a conservative estimate, $c$ would be unity. Using alpha-alpha, alpha-electron, and electron-electron anti-coincidence schemes (see below), we are able to reduce this value to $c=0.4$.

\emph{Signal to background ---} Combining Eqs.~(\ref{eqn:signal}) and (\ref{eqn:background}), we obtain the signal-to-background ratio
\begin{equation}
S/B(t) = \frac{C'_{IC}}{\epsilon_{\alpha} \cdot C'_{bg}} = \frac{b \cdot \eta \cdot \lambda \cdot \epsilon_{e^-} \cdot e^{-t/\tau}}{\epsilon_{\alpha} \cdot c \cdot A \cdot \tau},
\label{eqn:S/B}
\end{equation}
where $\epsilon_{\alpha}$ is the probability that the alpha particle gets into the Si detector and triggers the sweep. With the design considered here (a thin layer electroplated directly onto the Si detector), $\epsilon_{\alpha}=50\%$.

\emph{Signal to noise ---} In the experiment, we will average over time traces of some $10^7$ to $10^8$ U-233 alpha decays to accumulate sufficient statistics.

The number of sweeps $N_{sw}$ recorded over the total measurement time $T$ is $N_{sw}=T\cdot A \cdot \epsilon_{\alpha}$. The background is of purely Poissonian statistics, such that the noise of the measurement is simply the square root of the background, $N(t) = \sqrt{C'_{bg} \cdot N_{sw}}$, whereas the signal is $S(t)=C'_{IC} \cdot N_{sw}$. Using Eqs.~(\ref{eqn:signal}) and (\ref{eqn:background}), we obtain
\begin{equation}
S/N(t) = \sqrt{\frac{T \cdot \Delta t \cdot \epsilon_{\alpha}}{c}} \cdot  b \cdot \eta \cdot \lambda \cdot \epsilon_{e^-} \cdot \frac{e^{-t/\tau}}{\tau}
\label{eqn:S/N}
\end{equation}
for a given bin at time $t$.

Note that the signal-to-noise scales with measurement time and bin length in the familiar square-root behavior, but it is entirely independent on the sample's activity $A$. With all experimental parameters stated above, taking $\tau = 10\,\mu{\rm s}$, and looking at only a single characteristic time bin at $t=\tau$ after the alpha decay, we obtain a S/N of 1.0 for a measurement time of about two days ($T=1.9 \times 10^5$\,s). When sacrificing timing resolution and integrating over time bins of length $\tau \approx 10\,\mu{\rm s}$, the S/N takes a value of 4.5.

In conclusion, we have shown that the experiment described here is highly feasible. The \emph{a-priori} knowledge of the IC half-life in the stopping material of choice is of great advantage in the design of the measurement protocol. It shall be emphazised again that all parameters in Eq.~(\ref{eqn:S/N}) are well-known and fixed, or can only be varied within a very limited margin.

\section{Experimental setup}
\label{sec:expsetup}

\subsection{The samples}

\begin{figure*}
\centering
\includegraphics[width=\textwidth]{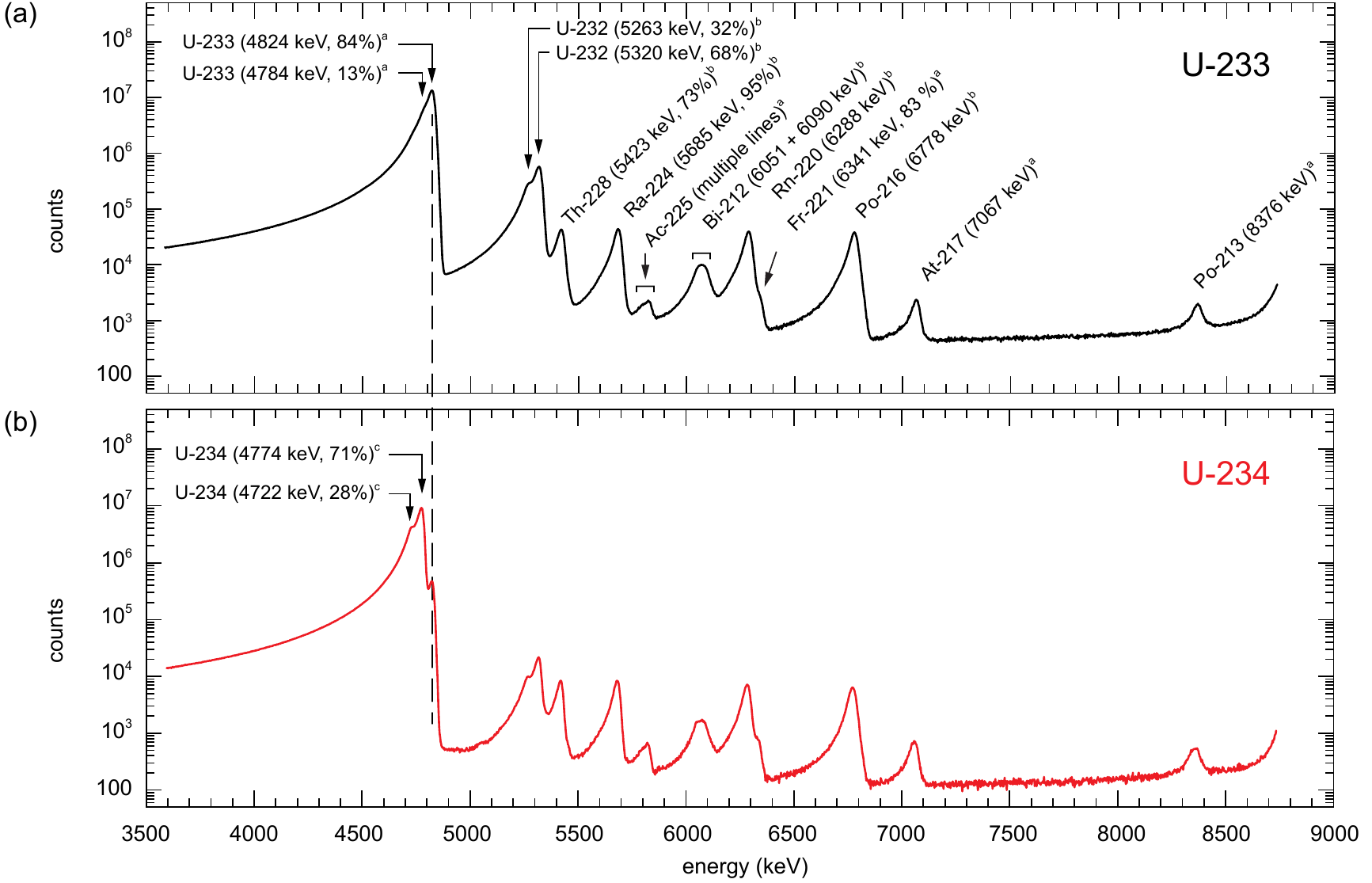}
\caption{Alpha spectra of (a) the U-233 sample, and (b) the U-234 sample. The thickness of the uranium layer is 1.6\,nm for U-233 and 1.1\,nm for U-234. In both cases, the sample is sandwiched in between 4\,nm of gold. The U-233 sample contains a 19.3\,ppm contamination of U-232 (4.5\% in activity), and the U-234 sample contains 1.4\% of U-233 contaminations. The decay chains are clearly visible, where superscripts ${}^{a, b, c}$ denote the U-233, U-232, and U-234 decay chains, respectively. 
}
\label{fig:fig3}
\end{figure*}

The U-233 material is attached directly onto a Si detector by molecular plating using 2-propanol as solvent \cite{Trautmann1989poa}. At first, the surface of the Si detector (CANBERRA Passivated Implanted Planar Silicon (PIPS) detector, active area: 600\,mm$^2$) is rendered conductive by plasma sputtering of a 4-nm gold layer. Thin films of Au tend to be brittle, therefore Pd is added to make the film softer and better attach to the Si surface. The detector then forms the basis of an electroplating cell with an inner diameter of 18\,mm. The U-233 material is dissolved with 1\,$\mu$L of 0.2\,M HNO$_3$, this solution is then filled into the electroplating cell together with the 2-propanol solvent. A platinum electrode is immersed in the solution, and a positive voltage is applied to the electrode. The current is set to 3\,mA, and the electroplating takes 60 minutes. Afterwards, the PIPS detector is coated with another 4-nm layer of gold to cover the uranium layer.

The alpha-particle spectrum of the U-233 sample is measured with a commercial alpha-particle spectrometer; see Fig.~\ref{fig:fig3}(a). The U-233 activity is $1420\,$Bq, and the thickness of the $^{233}$UO$_2$ layer is calculated to be 1.6\,nm.

We seek to compare U-233 data with U-234 data, as U-233 and U-234 are very similar in lifetime (160\,000\,a and 245\,000\,a, resp.) and energy $Q_{\alpha}$ released in the alpha decay (4824\,keV and 4772\,keV, resp.). Due to the finite instrumental resolution of our setup (about 100\,keV, limited by the electronics), it is not possible to prepare a mixture of U-233 and U-234 on a single PIPS detector and still resolve the individual components. Therefore, we prepare a U-234 sample on a second PIPS detector using the same protocol as described above. The U-234 activity is 630\,Bq, which corresponds to a $^{234}$UO$_2$ layer of 1.1\,nm thickness.

For the U-233 electroplating source, the activities of U-233, U-232, and Th-228 are 1420\,Bq, 63.7\,Bq, and 2.76\,Bq, respectively. Therefore, the amount of U-232 in the U-233 sample is 19.3\,ppm by mass and 4.5\% by activity. The time since chemical purification is about 90 days. Note that electroplating of thorium is a factor of about 2 times less efficient compared to uranium.

For the U-234 electroplating source, the activities of U234, U-233, U-232, and Th-228 are 626\,Bq, 13.3\,Bq, 1.57\,Bq, and 0.328\,Bq, respectively. Therefore, the contamination of U-233 in the U-234 1.41\% by mass, and the time since purification is 236 days.

\subsection{The setup}

The experimental setup has been shown in Fig.~\ref{fig:fig1}(a). The uranium sample on the Si detector is placed inside a vacuum chamber evacuated to better than $10^{-6}\,$mbar. Three Ni electroformed meshes (single transparency: 88\%) are placed above the sample, a metal mesh (transparency: 80\%) is placed above the meshes. A 2-stage MCP detector (HAMAMATSU F 1942-04G, effective area diameter 77\,mm) is placed above the mesh. The distance between the sample and the MCP detector is 17\,mm, and the solid angle coverage of the upper hemisphere is about 70\%. Combining the total transmission through the grids with the solid angle yields a geometric efficiency of $\lambda \approx 40\%$ . A voltage of $+500\,$V is applied to the surface of the MCP detector in order to maximize the detection efficiency for electrons ($\epsilon_{e^-} \approx 50\%$).

The Ni mesh in the lowest position and the mesh in the highest position are grounded, and a negative voltage (between 0 and $-6\,$V) is applied to the other two Ni meshes as a retarding field to analyze the energy of the electrons from the uranium sample.

The charges produced by the alpha particle inside the Si detector are collected into a charge-sensitive preamplifier (MESYTEC MPR-1). The output from the preamplifier is sent to a pulse shaping amplifier (ORTEC 570), and the timing filter output of the preamplifier is sent to a discriminator (ORTEC 584). The output from the pulse shaping amplifier and that from the discriminator are sent to a 16-input PHA \& LIST module (NIKI GLASS A3100), and then the pulse height information (\textit{i.e.}, the energy information) and the timing information of the alpha-particle signals are registered by a PC.

The output from the MCP detector is sent to an amplifier (HAMAMATSU C5594), and its output is fed into a discriminator (ORTEC 584). The output from the discriminator is sent to the 16-input PHA \& LIST module mentioned above, and the timing information of the MCP signal is registered by the PC. Whenever there are two signals from the MCP detector within a time window of $1.1\,\mu$s, a pulse is produced with the help of the coincidence electronics and  sent to the 16-input PHA \& LIST module. The timing information of such multi-electron signals within $1.1\,\mu$s is registered by the PC as well.
 
The time stamps of the Si detector (alpha particles) and the MCP (electrons) are used to construct the time trace of electrons following an alpha decay. Therefore, every recorded alpha decay is used as a trigger, and we record all MCP signals in the following $100\,\mu$s with a resolution of 500\,ns (``sweep''). Whenever there is a second alpha decay within this span of $100\,\mu$s, the data is excluded. Two or more MCP signals registered within $1.1\,\mu$s are likely prompt electrons from an alpha decay, and not a signature of the isomer decay, which would result in only a single electron. Therefore, such events are also excluded to reduce the background.

A typical time trace is shown in Fig.~\ref{fig:fig1}(b). Less than one count is registered per alpha decay, and we typically average over $10^8$ sweeps to build up sufficient statistics. Following the alpha decay, we observe a prompt signal on the MCP with an amplitude of about 0.1 counts and a FWHM duration below 500\,ns, followed by an exponential decay with a constant of about $5\,\mu$s.

\begin{figure}
\includegraphics[width=\columnwidth]{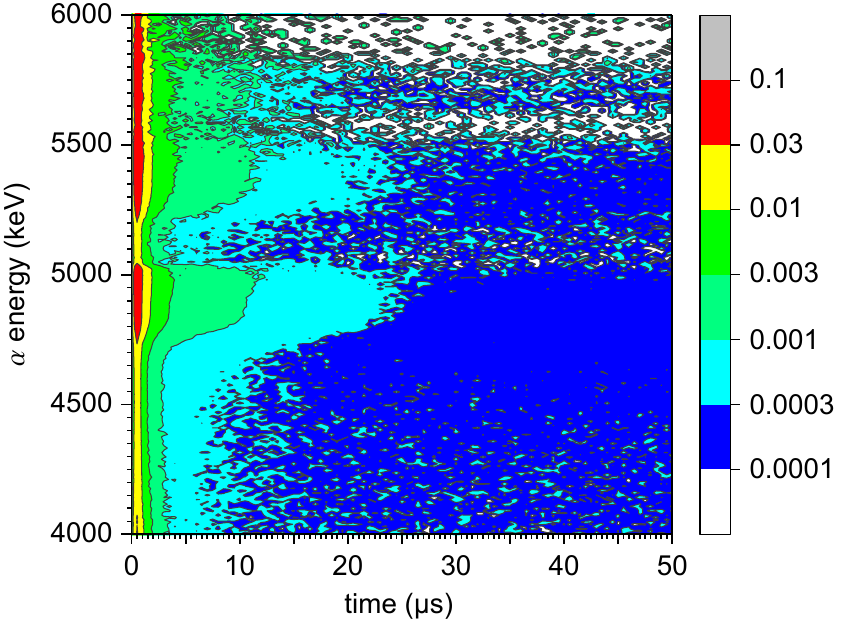}
\caption{Time dependence of electron emission registered by the MCP, in dependence of the energy of the alpha particle that triggered the detection. For this data set, the U-233 sample was used, and the retarding field was set to 0\,V. The signal is recorded in time bins of 500\,ns, and the energy is recorded in bins of 0.86\,keV width. The signal at about 4800\,keV (5300\,keV) corresponds to the decay of U-233 (U-232). The MCP counts are normalized to the number of sweeps.}
\label{fig:fig4}
\end{figure}

Together with the time stamp of each alpha decay, we also record the energy of the alpha particle in bins of 0.86\,keV width. This allows us to analyze the electron time trace for each type of alpha decay (uranium and all its daughters) individually, \textit{e.g.}~comparing U-233 to U-232. A representation of such a data set is shown in Fig.~\ref{fig:fig4}.

For one dataset, we typically sum over $2\times10^8$ alpha decays, which takes about three days. Such datasets are obtained for eight different retarding voltages between 0 and $-6\,$V. After taking these eight datasets with the U-233 sample, we open the vacuum chamber, exchange the Si detector covered with U-233 for another one covered with U-234, and take the reference data set under identical conditions.

\section{Results}

\subsection{Time traces}

Time traces of the MCP signal are initiated by a detected alpha decay and recorded for $100\,\mu$s. The large prompt peak is not of interest here, as it bears no information on the isomer. Therefore, we only analyze the signal between 10 and 100\,$\mu$s.

The lines of the alpha spectrum corresponding to the vertical axis of Fig.~\ref{fig:fig4} are broadened considerably by the passage of the alpha particles through the uranium (1.6\,nm), gold (4\,nm), and silicon (50\,nm) layers into the PIPS detector at all angles. The larger the angle from normal incidence, the lower the recorded energy of the alpha particle. In our analysis, we take into account only data for alpha energies between 0 and 250\,keV below the nominal value of $Q_{\alpha}$, which corresponds to angles of up to $87.5^{\circ}$ from normal incidence. This range contains 87\% of all alpha particles registered by the detector.

\begin{figure}
\includegraphics[width=\columnwidth]{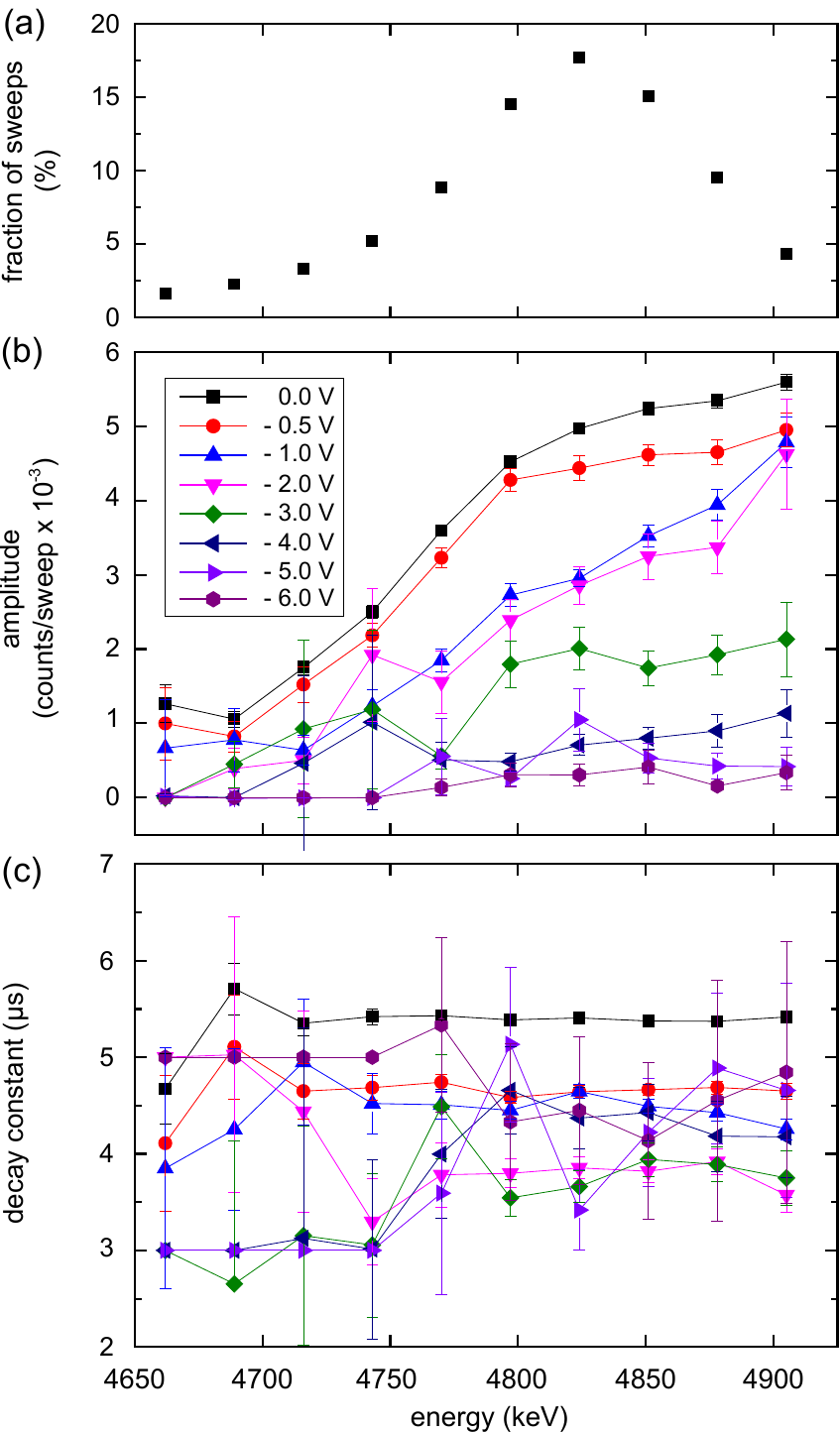}
\caption{Fitting a single exponential decay to the U-233 data, where the dataset is subdivided into bins of 25\,keV width. (a) Fraction of the total number of sweeps in each energy bin. (b) Amplitude and (c) decay constant of the exponential fit to the data; lines are only a guide to the eye.}
\label{fig:fig5}
\end{figure}

The temporal evolution of the signal for $t>10\,\mu$s can be described exceptionally well by a single exponential decay plus an offset (corrected $R^2 > 0.995$). We subdivide the U-233 data into bins of 25\,keV width and fit a single exponential function to each subset. The amplitudes and $1/e$ time constants are shown in Fig.~\ref{fig:fig5}. The amplitude of the detected electrons decreases with increased retarding voltage, where the decrease can be approximated by an exponential dependence (decay constant $1.5(5)/\mathrm{V}$). At our highest retarding voltage of $-6\,$V, already 96\% of all electrons are rejected.

The amplitude also decreases monotonically for larger incident angles of the alpha particle, evidenced by a smaller detected alpha energy. As the alpha particle and the recoil ion are ejected in opposite directions, larger incident angles of the alpha particle correspond to a larger average implantation depth of the recoil nucleus, and thus a smaller probability of generating low-energy electrons that leave the sample.

The offset is entirely flat for $t>50\,\mu$s, it is formed by uncorrelated radioactivity and shows purely Poissonian statistics. This background is only weakly dependent on the retarding voltage and amounts to $C'_{bg}=3 \times 10^{-4}$ for bin lengths $\Delta t = 500\,{\rm ns}$ and an activity of $A=1420\,$Bq. The offset is entirely independent of the alpha energy ($Q_{\alpha}$ between 4 and 8\,MeV), indicating again that this background is caused by uncorrelated radioactivity. 

The exponential decay constant is independent of the implantation depth of the recoil ion and depends mildly on the retarding voltage; see Fig.~\ref{fig:fig5}(c). We deduce a $1/e$ decay constant of $5(1)\mu$s.

Both the energy distribution of the delayed electrons (median energy $2-3\,$eV) and their decay constant are compatible with the sought-after IC electron from the de-excitation of Th-229m. Their overall amplitude, however, is 3 orders of magnitude larger than the IC electron signal estimated in Eq.~(\ref{eqn:signal}). All parameters that enter into Eq.~(\ref{eqn:signal}) have been determined experimentally (such as the 2\% branching ratio and the 7-$\mu$s half-life), have been simulated carefully (such as the efficiency $\eta \approx 1.8\%$), or are given by the experimental setup (MCP detection efficiency ($\epsilon_{e^-} \approx 50\%$ and geometric transmission through the grids $\lambda \approx 40\%$), such that our estimate will not be wrong by orders of magnitude. As a consequence, we conclude that the observed signal \emph{cannot} be explained by IC electrons. The discrepancy between observed and detected signal is shown in Fig.~\ref{fig:fig6}.

The amplitude of the observed low-energy electrons increases with $Q_{\alpha}$. We speculate that the recoil nucleus creates lattice defects in the stopping layer, with low-energy electrons being produced during lattice relaxation. This process has been described as exoelectron emission \cite{Claverie2004sfn}. We suspect that the Si material of the PIPS detector could also be source of low-energy electrons. These low-energy electrons mask the IC electrons from the de-excitation of Th-229m.

\subsection{Comparison between U-233 and U-234}

\begin{figure}
\includegraphics[width=\columnwidth]{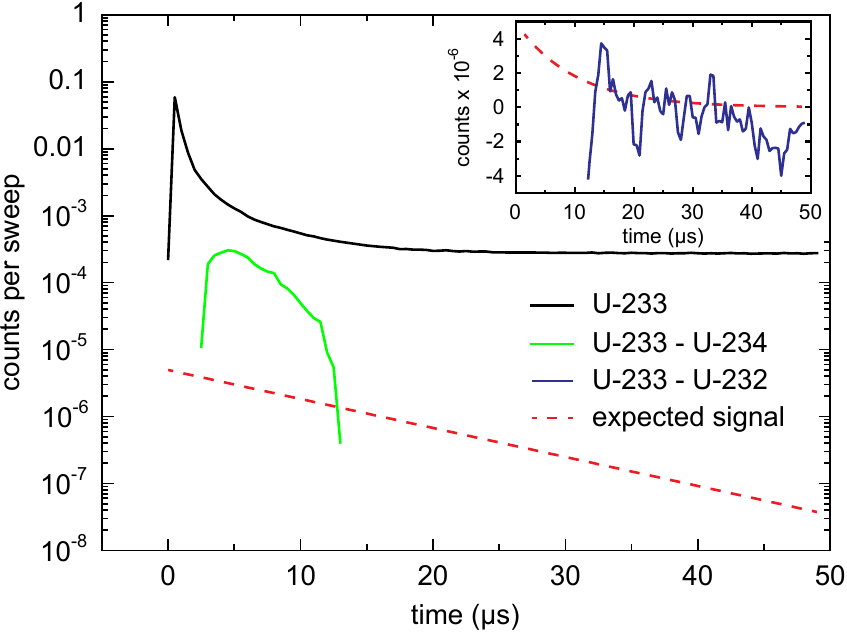}
\caption{The observed signal for U-233 (solid black line) is orders of magnitude stronger than the expected isomer signal (dashed red line). Also the difference signal between measurements with U-233 and U-234 is considerably larger than the expected signal (green). Inset: Difference signal between U-233 and U-232; see the text for details.}
\label{fig:fig6}
\end{figure}

To tease out the contribution of IC electrons to the observed signal of delayed low-energy electrons, we perform a reference measurement with U-234. The difference in alpha energies of U-233 and U-234 is only 50\,keV. The emission of low-energy electrons from the sample surface depends on the chemical composition, but it is not expected to be isotope-specific. With comparable half-lives (245\,000\,a and 159\,000\,a), U-234 appears to be an ideal reference system for U-233.

The difference in $Q_{\alpha}$ is too small to be resolved in our experiment. In addition, the strongest line in U-234 is within 10\,keV of the second line in U-233: these two isotopes cannot be measured within the same sample. We therefore perform a set of reference measurements under identical conditions with a U-234 sample prepared under the same conditions as the U-233 sample.

Data analysis is performed as with the U-233 sample. The amplitudes of the delayed electron signal are 20(2)\% lower compared to the U-233 sample, and there is a slightly longer decay constant of $6(1)\,\mu$s. This is a detrimental finding for our strategy of comparing U-233 and U-234 to single out the contribution of the Th-229m IC electron:~the emission of low-energy electrons seems to be sensitive to the sample fabrication up to a degree that cannot be controlled in our experiment (\textit{e.g.}~layer thickness, surface roughness and contaminations).

The difference of the U-233 and U-234 signals is shown in Fig.~\ref{fig:fig6}. The contribution of the IC electrons would show up as an exponentially decreasing excess in the U-233 signal compared to U-234. Such a feature is found here, but it is at least two oders of magnitude larger than expected:~we conclude that this excess cannot be identified with the Th-229m isomer de-excitation. To avoid the stark systematic effect of variations in the structure of the stopping layer, future measurements shall seek to compare two different isotopes within a single sample.

\subsection{Comparison between U-233 and U-232}

\begin{figure}
\includegraphics[width=\columnwidth]{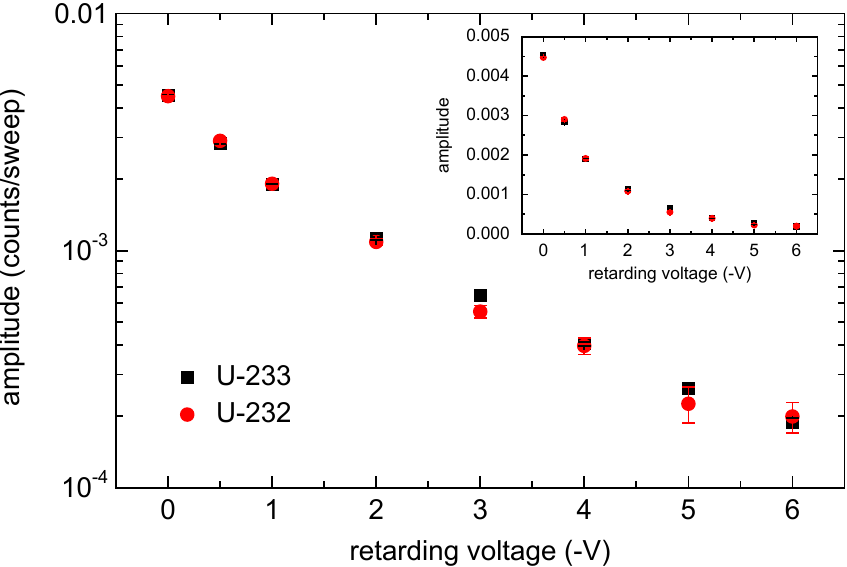}
\caption{Comparison of the exponential decay of the electron signal for U-233 (black squares) and U-232 (red dots). We plot the amplitudes of the single exponentials fitted to the data:~within the error bars, they are identical for the two isotopes. Inset: Same data on a linear scale. }
\label{fig:fig7}
\end{figure}

In a next step, we compare the U-233 data to data of U-232 (half-live 68.9\,a), which appears as a 20\,ppm contamination in our U-233 material. The U-232 activity is 4.5(1)\% the activity of the U-233 content.  Again, we fit single-exponential decay functions to the data for $t>10\,\mu$s. The amplitudes are considerable larger than for the U-233 data, as expected for the larger energy released in the U-232 alpha decay ($\Delta Q_{\alpha}=500\,$keV). The amplitude of low-energy electrons is 11(1)\% larger compared to U-233, which corresponds approximately to the relative difference in $Q_{\alpha}$. For U-232, we obtain decay constants that are within the error bars (less than 2\% deviation) of the values obtained for U-233.

The difference between the U-233 and the U-232 data sets for zero retarding voltage is shown in the inset of Fig.~\ref{fig:fig6}. Here, the U-232 data has been scaled by a factor of 0.87 to compensate for the difference in $Q_{\alpha}$, and averaged over $\tau/2 = 5\,\mu$s to reduce the noise. Clearly, the noise is too large to identify the contribution of the isomeric signal. The noise originates predominantly from the U-232 dataset, as the U-232 activity of roughly 20 times smaller than the U-233 activity. Similar activities for both isotopes, and possibly an increased measurement time, should be sufficient means to reduce the noise to a level that allows for an unambiguous identification of the isomeric signal. This is the main finding of the study presented here.

The amplitudes of exponential fits to the data for $t>10\,\mu$s are shown in Fig.~\ref{fig:fig7} for various retarding voltages. Here, the decay constant was fixed to $5.2\,\mu$s for all decay curves. Aside from a scaling with a constant factor to compensate for $\Delta Q_{\alpha}$, we find no evidence for a dependence of the electrons' energy spectrum on the energy $Q_{\alpha}$ released in the alpha decay. This is an important finding for a future energy measurement of the IC electron. The sought-after signal would then show up as an additional component only in the U-233 data, with a dependence on the retarding voltage derived in Ref.~\cite{Seiferle2017fso}.

\subsection{Variation of the stopping layer material}

In the study presented thus far, we have identified the delayed low-energy electrons as the main roadblock in the detection of IC electrons. The amplitude, energy distribution, and decay constant of these electrons depends on the surface material. So far, we have used Au as the stopping layer, as the IC half-life of Th-229m in Au had already been measured independently. Now, we will try other stopping materials. As the amplitude of low-energy electrons depends critically on the sample fabrication (\textit{e.g.}~layer thickness and surface roughness), we will focus on the decay constant.

\emph{Uranium} --- In a first attempt, we increase the U-233 layer thickness and leave away the Au top layer: the Th-229(m) recoils are now stopped directly in the U-233 material. The decay constant of low-energy electrons is 12(2)\,$\mu$s, as compared to 5(1)\,$\mu$s in Au, at a similar amplitude, which clearly is a disadvantage. Variation of the temperature might prove to be beneficial, but has not been part of this study.

\emph{Carbon foil} --- In an alteration of the setup, we deposit a thin layer of U-233 onto a carbon foil of 10\,nm thickness. The foil is placed in between the PIPS detector (below) and the MCP (above). The U-233 layer faces the PIPS detector, such that Th-229(m) recoils are stopped in the foil. In this setup, we observe the same prompt peak as with a metallic stopping layer, but no delayed signal at all: the prompt signal merges directly into the flat background at 4(1)\,$\mu$s after the alpha decay, even for zero retarding voltage. We conclude that no low-energy electrons are expelled from the carbon foil, making this a promising candidate for future studies. This experiment was performed with a sample of 3\,mm diameter:~future work will investigate the mechanical stability of cm-sized samples.

\section{Conclusion}

We have presented a new attempt to measure the energy of the Th-229 isomer via spectroscopy of the IC electrons. The experiment is very compact and simple in design, it builds on the measurement of low-energy electrons in coincidence with U-233 alpha decays inside a stopping layer. The stopping layer is designed to optimize the combined yield of stopping the Th-229m recoil and allowing the IC electron to leave the material.

The sought-after signature of the IC electron is masked by the emission of electrons from the sample surface, which are very similar to the IC electron in energy and decay time constant, yet more numerous than the expected IC electron signal by 3 orders of magnitude. The number of these background electrons depends critically on the surface properties and cannot be controlled at the level of the expected IC electron signal.

For future experiments, we suggest to reference the U-233 signal to the signal of U-232, where the difference in decay energy $Q_{\alpha}$ of these two isotopes is larger than the detector resolution. These two isotopes can thus be measured within the same sample, such that they probe the same surface properties to reduce systematic effects. Owing to its comparably short half-life, addition of a small quantity (440\,ppm) of U-232 to the U-233 sample already yields the same activity. An excess component in the U-233 signal would be interpreted as the contribution by the IC electron.

Further, we suggest to continue the assessment of various stopping layer materials to minimize the amplitude, energy, and decay time constant of the emitted electrons. Best results in this respect have been obtained with thin diamond-like carbon foils.

\section*{Acknowledgements}

We thank Lars v.~d.~Wense, Benedict Seiferle, Peter G. Thirolf, Andreas Fleischmann, Piet van Duppen, and Jan Welch for fruitful discussions. The U-234 material was generously provided by the LMU Munich group of Peter G. Thirolf. Further, we thank Dieter Heinz, Hanno Filter, and Johannes Sterba of the Atominstitut in Vienna for technical support and loan of equipment, as well as Jakob Gruber for the plasma sputtering. We are grateful to T.~Munakata and T.~Yamada for advice and support in the development of the retarding-field electron analyzer. This project has received funding from the European Union's Horizon 2020 research and innovation programme under grant agreement No.~664732  (``nuClock'') and was supported by the Austrian Science Foundation FWF under grant No.~F41 (SFB ``VICOM''), by the WWTF project No.~MA16-066 (``SEQUEX''), and by JSPS KAKENHI Grant No.~25800150 and 16J06700.

\appendix

\section{Estimation of parameter $\eta$}
\label{app:I}

Here, we describe the semi-analytical method for the estimation of the parameter $\eta$ from Eq.~(\ref{eqn:signal}), \textit{i.e.}, the probability that after the decay ${\rm ^{233}U} \rightarrow {\rm ^{229(m)}Th}$, the alpha particle will be emitted towards the PIPS detector, the Th-229(m) recoil nucleus will be stopped either within the uranium source or within the gold stopping layer, and that the conversion electron will leave this layer towards the MCP. Such an estimation is relevant for the optimal choice of the thickness of the stopping layer:~it should be thick enough to stop recoil nuclei, but thin enough to allow conversion electrons to leave into free space.
\begin{figure}
\centering
\includegraphics[width=\columnwidth]{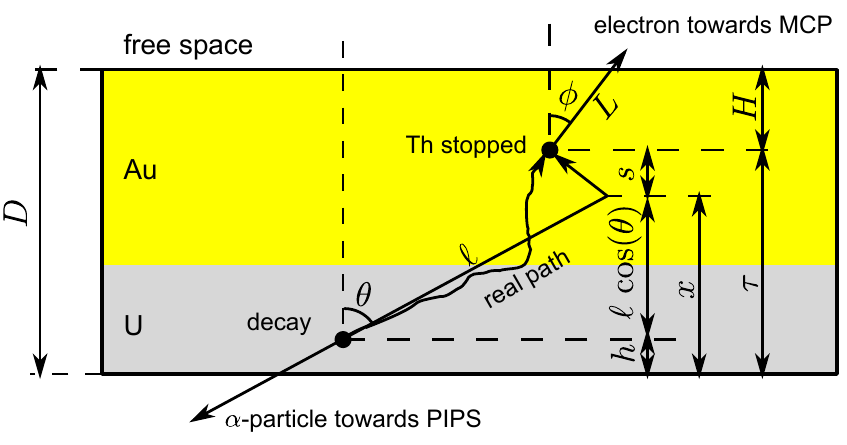}
\caption{Schematic drawing of the paths of the alpha particle, the recoil nucleus, and the conversion electron.}
\label{fig:paths}
\end{figure}

The Th-229(m) recoil nuclei have an energy of about 84~keV. We suppose that the alpha particle is registered by the PIPS detector, if and only if it is emitted into the lower hemisphere, see Fig.~\ref{fig:paths}. We represent the total vector transition of the recoil nucleus in the form
\begin{equation}
\vec{r}-\vec{r}_0=\vec{\ell}+\vec{S},
\label{eq:A1}
\end{equation}
where $\vec{\ell}$ is a vector with fixed length $\ell$ (projected range), whose direction is randomly distributed within the upper hemisphere. $\vec{S}$ is a random vector with zero mean and normally distributed components. We simplify the geometry by supposing that the components of $\vec{S}$ have an isotropic normal distribution, and their dispersion $\sigma$ is equal to the arithmetic mean of longitudinal $\sigma_\ell$ and lateral $\sigma_{\perp}$ stragglings for 84-keV Th-229 nuclei in gold given by the SRIM code \cite{SRIM}. For gold, $\sigma_\ell = 5.42$\,nm and $\sigma_\perp = 3.92$\,nm; we average to $\sigma=4.7$\,nm and $\ell=9.3$\,{nm}. The values for uranium are very similar, $\sigma_\ell = 6.36$, $\sigma_\perp = 4.52$, and $\ell=9.9$\,{nm}. Such an approximation simplifies the analysis significantly, but does not lead to critical errors in the optimization of the stopping layer thickness.

We are interested only in the final depth $H$, which can be written as:
\begin{equation}
H=D-h-\ell \cos(\theta)-s=D-h-\ell \chi - s. \label{eq:A2}
\end{equation}
Here, $h$ is the random intitial position of the nucleus within the uranium layer, $\ell$ is the deterministic projected range, $\chi=\cos(\theta)$ is the random cosine of the polar angle, and $s$ is the vertical component of the random vector $\vec{S}$. The probability distributions of $h$, $\chi$, and $s$ are:
\begin{align}
p_h(h) =& \left\{
\begin{array}{l}
\frac{1}{D_U},\, h \in [0,D_{\rm U}] \\
0, \, h \in (-\infty,0) \cup (D_{\rm U}, \infty)
\end{array}
\right. \label{eq:A3}
\\
p_\chi(\chi) =& \left\{
\begin{array}{l}
\frac{1}{2}, \, \chi \in [0,1] \\
0, \, \chi \in (-\infty,0) \cup (1, \infty)
\end{array}
\right. \label{eq:A4}
\\
p_s(s) =& \frac{\exp\left(-\frac{s^2}{2 \sigma^2}\right)}{\sqrt{2 \pi \sigma^2}}.
\label{eq:A5}
\end{align}
Here, $p_\chi$ is normalized to $1/2$, because we consider only the nuclei emitted into the upper hemisphere. Representing
\begin{align}
H&=D-x-s\quad {\rm with } \quad x=h+\ell \chi, \label{eq:A6}
\end{align}
where $D=D_{\rm U}+D_{\rm Au}$ is the sum of thicknesses of the uranium source and the gold stopping layers, and expressing
\begin{align}
p_x(x) & = \int\limits_{0}^{D_{\rm U}} p_h(h)p_\chi\left(\frac{x-h}{\ell}\right) \frac{dh}{\ell}
= \frac{1}{2D_{\rm U} \ell}
\label{eq:A7} \\
& \times
\left\{ 
\begin{array}{ccl}
0 &,& x<0; \\
x &,& 0<x<\min(D_{\rm U},\ell) \\
\min(D_{\rm U},\ell) &,& \min(D_{\rm U},\ell) < x < \max(D_{\rm U}, \ell) \\
\ell+D_{\rm U}-x &,& \max(D_{\rm U},\ell)<x<D_{\rm U}+\ell \\
0 &,& D_{\rm U}+\ell<x
\end{array}
 \right.
 \nonumber \\
p_H(H) & = \int\limits_{0}^{\infty} p_x(x)p_s (D - H- x) dx,
\label{eq:A8}
\end{align}
we can find the probability distribution of the depths of the thorium nuclei recoiled into the upper hemisphere.

Substituting Eq.~(\ref{eq:A8}) into (\ref{eq:eta}), we can estimate the probability $\eta$. Probabilities evaluated according to this method are presented in Fig.~{\ref{fig:fig1}\,(b)} by styled colored curves. One may see that within the broad range of inelastic mean free paths $\lambda_e$, the optimal thickness of the gold stopping layer is about 5~nm. Comparing the estimated probabilities with the one obtained with the help of the SRIM simulation, we find that the approximate method overestimates $\eta$ for short electron ranges. This can be explained by the increased probability of a nuclei stopped in close vicinity of the surface to leave the stopping layer during the random walk phase at the end of its path.

\bibliographystyle{apsrev}
\bibliography{bib}

\end{document}